\documentclass{article}

\usepackage{arxiv}

\usepackage[utf8]{inputenc} % allow utf-8 input
\usepackage[T1]{fontenc}    % use 8-bit T1 fonts
\usepackage{hyperref}       % hyperlinks
\usepackage{url}            % simple URL typesetting
\usepackage{booktabs}       % professional-quality tables
\usepackage{amsfonts}       % blackboard math symbols
\usepackage{nicefrac}       % compact symbols for 1/2, etc.
\usepackage{microtype}      % microtypography
\usepackage{lipsum}

\usepackage[english]{babel}
\usepackage{multicol}
\usepackage{color,graphicx,amsmath,amssymb}
\usepackage{eurosym}
\usepackage{fancyhdr}

\usepackage{hyperref}
\usepackage{caption}
\usepackage{subcaption}
\usepackage{multirow}
\usepackage[table,xcdraw]{xcolor}

\usepackage{graphicx}
\usepackage{tikz}
\usetikzlibrary{positioning}
\usetikzlibrary{arrows.meta}
\usetikzlibrary{decorations.pathreplacing}
\usetikzlibrary{calc}
\usetikzlibrary{shapes}

\definecolor{UCAOrange}{RGB}{210, 115, 45} % (82, 45, 18)
\definecolor{UCABlue}{RGB}{5, 100, 135} % (2, 39, 53)

\newcommand{\Nt}{N_{\rm train}}
\title{NILM as a regression versus classification problem: the importance of thresholding}

\author{
  Daniel Precioso\\%\thanks{Use footnote for providing further information about author (webpage, alternative address)---\emph{not} for acknowledging funding agencies.} \\
  Department of Computer Science\\
  Higher School of Engineering\\
  Universidad de C\'adiz, Spain\\
  \texttt{daniel.precioso@uca.es} \\
   \And
 David G\'omez-Ullate\thanks{On leave of absence from Department of Theoretical Physics, Universidad Complutense de Madrid, Spain.}\\
  Department of Computer Science\\
  Higher School of Engineering\\
  Universidad de C\'adiz, Spain\\
  \texttt{david.gomezullate@uca.es} \\
}

\begin{document}
\maketitle

\begin{abstract}
Non-Intrusive Load Monitoring (NILM)  aims to predict the status or consumption of  domestic appliances in a household only by knowing the aggregated power load. NILM can be formulated as regression problem or most often as a classification problem. Most datasets gathered by smart meters allow to  define naturally a regression problem, but the corresponding classification problem  is a derived one, since it requires a conversion from the power signal to the status of each device by a thresholding method. We treat three different thresholding methods to perform this task, discussing their differences on various devices from the UK-DALE dataset. We analyze the performance of deep learning state-of-the-art architectures on both the regression and classification problems, introducing criteria to select the most convenient thresholding method.
\end{abstract}

% keywords can be removed
\keywords{Non Intrusive Load Monitoring \and Time series \and Convolutional Neural Networks \and Recurrent Neural Network}

\section{Introduction}

Non-Intrusive Load Monitoring (NILM) was proposed on 1992 by G. W. Hart~\cite{hart1992nonintrusive} as a method to predict whether an individual appliance is active or inactive, by observing the total aggregated power load, having information of the nominal consumption of each appliance in each state. The first approach to NILM employed Combinatorial Optimization, which at the time was the standard technique for disaggregation problems. For historical review of the evolution of NILM techniques, see for instance \cite{nalmpantis2018review, do2016applications, klemenjak2016nonintrusive}. This first approach had a major shortcoming: Combinatorial Optimization performed the power disaggregation on each instant independently of the others, without considering the load evolution through time. These original algorithms were very sensitive to noise and were only accurate for houses with very few appliances. Thus, they could not be applied to real-life scenarios.

NILM algorithms received renewed attention at beginning of the 21st century, mostly thanks to the increased number of datasets coming from smart electric meters installed in domestic residences. These meters are able to record the power load from a household at short time intervals (one hour or less) and send those values to the electric company~\cite{smartmeter}.  The first open-source NILM datasets were published in 2011 and they triggered further research activity by setting benchmarks for comparison of models. These datasets stored high frequency time series for both aggregated and appliance power load~\cite{kolter2011redd, anderson2012blued, kelly2015ukdale}. For a recent comprehensive review on NILM datasets, see~\cite{klemenjak2019datasets, klemenjak2020comparability}. Many of them are publicly available in the NILM-EU wiki.\footnote{\href{http://wiki.nilm.eu/datasets.html}
{http://wiki.nilm.eu/datasets.html}}

Soon after these datasets became available, the prevailing approach to NILM shifted from the combinatorial optimization problem mentioned above, to a supervised learning problem with times series in Machine Learning ~\cite{revuelta2017survey, faustine2017survey}. Traditional ML methods such as Hidden Markov Models were initially used~\cite{kim2011hmm, kolter2012hmm, jia2015nfhmm, mengitsu2018hmm}, while in recent years the incredible growth of deep learning algorithms has dominated the field~\cite{kelly2016disaggregation, Kim2017NonintrusiveLM, krystalakos2018sliding, harell2019wavenilm, kasemili2019bilstm, kyrkou2019timeseries, massidda2020tpnilm}.

Much of the recent effort of the NILM research community has been focused on improving the efficiency of prediction algorithms and computational speed, usually by presenting novel model architectures~\cite{kyrstalakos2018sliding, kyrkou2019timeseries} or even trying unconventional approaches such as  finite state machines~\cite{ducange2014fuzzy} and hybrid programming~\cite{kong2016hybrid}.

However, fewer works are devoted to investigating the formulation of the problem, and how this can affect the algorithm's performance. For instance, uniformization of evaluation metrics for easy comparison of different models has been stressed in \cite{pereira2018metrics,makonin2015nonintrusive}, while the dependence on sampling frequency has been treated in~\cite{ruano2019commercial}. This becomes a relevant issue when trying to leverage theoretical NILM results on high frequency benchmark datasets to real life scenarios where records are sampled at lower rates.

NILM datasets typically include both the aggregated power load and that of each monitored device, but not the device status (i.e. whether it is ON or OFF). Thus, a regression problem to predict the consumption of each device is naturally defined by the data. However, most works in NILM address the classification problem of determining whether the device is ON or OFF, rather than its consumption at each time interval. Defining a classification problem requires establishing a threshold or some procedure to determine the output categorical variable from the continuous output power load. Our main observation is that this process involves an external choice of thresholding method which is not included in the initial problem formulation. Depending on how this preprocessing step is performed, the performance and interpretation of the final results may vary in a significant manner. The main contribution of this paper is to highlight this matter and to discuss several possible ways to define a classification problem from the native regression problem.

The paper is organized as follows: in Section~\ref{sec:problem} we formally introduce the necessary notation to define regression and classification problems for time series in supervised learning. Section~\ref{sec:threshold} introduces three different thresholding methods. Two different deep learning models are introduced and explained in Section~\ref{sec:models}. The purpose of studying two different models is to have more robust results and to ensure that the reported variations are not model dependent.
The detailed methodology is carefully explained in Section~\ref{sec:methodology}, including data preprocessing, definition of training, validation and test sets, loss functions, optimization algorithms, and evaluation metrics. The results are exhibited in Section~\ref{sec:results} for three monitored devices with different characteristics. Together with the results, we include a discussion on the criteria to choose the most convenient thresholding method. In section~\ref{sec:weights} we introduce architectures that optimize both classification and regression, to see how the two formulations interplay. Finally, we gather concluding remarks and outline  open research problems in Section~\ref{sec:outlook}.

%%%%%%%%%%%%%%%%%%%%%%%%%%%%%%%%%%5
\section{Problem Formulation}\label{sec:problem}
%%%%%%%%%%%%%%%%%%%%%%%%%%%%%%%%%5
NILM is formulated as a supervised learning problem, where the model is trained to take the aggregate power as input signal and predict the power or state (ON/OFF) of each monitored appliance.  This power load is measured by the smart meter at a constant rate $\tau^*$, which produces a series of power measurements\footnote{Strictly speaking, smart meters measure the energy consumption during the interval $(t,t+\tau^*)$ divided by the length of the interval $\tau^*$.} $P_i$ at each sampled interval.
For the analysis, it is often convenient to resample the original series at a larger sampling interval $\tau$, which is part of the preprocessing step. For instance, in this paper the native sampling interval for the UK-DALE dataset provided by the meters is $\tau^*=6s$, but we choose to resample the series at intervals of $\tau=60s.$

The aggregate power $P_j$ at instant $j$ is the sum over all appliances:

\begin{equation}
    P_{j} = \sum^{L}_{\ell=1} P_{j}^{(\ell)} + e_{j},
    \label{eq:aggregate_load}
\end{equation}
where $L$ is the total number of appliances in the building, $P^{(\ell)}_{j}$ is the power of appliance $\ell$ at time $j$, and $e_{j}$ is the unidentified residual load. All of these quantities are expressed in watts.

After resampling, the training set comprises a sequence of $n_{tot}$ records that we label as $\{P_i\}_{i=0}^{n_{tot}}$. This series is split in chunks of size $n$ that we group in vectors as  $\mathbf{P}_j=(P_{jn},P_{jn+1},\dots, P_{jn+n-1})$. We have a total of $n_{train}=n_{tot}/n$ such series, each of which will be an input to the model. The output of the model are sequences $\mathbf{P}_j^{(\ell)}$ for each monitored appliance over the same time intervals. The pairs $\{ \mathbf{P}_j, \mathbf{P}_j^{(\ell)} \}_{j=1}^{n_{train}}$ are considered as independent points in the training set. 

Supervised learning problems are usually referred to as classification or regression problems depending on whether the output variables are categorical or continuous. In the NILM literature, both of these approaches have been considered in different contributions, but there has been hardly no works devoted to the interplay between both formulations. It is precisely this gap that we would like to fill with this analysis.

In the \textit{regression approach}, the predicted quantities are the power load $\mathbf{P}_j^{(\ell)}$ for each device.

In the \textit{classification approach} the focus is on predicting whether a given appliance is at time $j$ in a number of possible states, typically ON or OFF. We assume therefore, for the sake of simplicity, that the appliance $\ell$ can be in one of two states at time $j$, which are $s^{(\ell)}_{j}=0$ (OFF state) and $s^{(\ell)}_{j}=1$ (ON state). It is not evident to ascertain when a given appliance is ON or OFF by just looking at the power load. Thus, the usual criterion is to establish a threshold $\lambda^{(\ell)}$ for each appliance, and define
\begin{equation}
    s_{j}^{(\ell)} = I(P_{j}^{(\ell)} \geq \lambda^{(\ell)}).
    \label{eq:status}
\end{equation}

Multi-state classification problems, where appliances may have more than one ON status (each of them  with a different consumption) have also been considered in the literature~\cite{desai2019multistate}. 
A correct definition of the classification approach thus involves a choice of threshold $\lambda^{(\ell)}$ for each appliance $\ell$. Ideally, this threshold should be determined by the series of data $\mathbf{P}_j^{(\ell)}$ alone, rather than being externally fixed by human intervention. In this paper we review different algorithms to determine this threshold, which lead to rather different outcomes depending on the complexity of the input signal. We address these methods in the following section.

%To alleviate the equations to come, we will omit the supper-index $(\ell)$. Reader should always assume that every value refers to only one single appliance $\ell$ at a time.

%%%%%%%%%%%%%%%%%%%%%%%%%%%%%%%%%%%%%%%%%5
\section{Thresholding}\label{sec:threshold}
%%%%%%%%%%%%%%%%%%%%%%%%%%%%%%%%%%%%%%%%%%%5

In this section we will explore different methods on how to set a threshold to determine the OFF and ON status of an appliance, given its input power signal.

\subsection{Middle-Point Thresholding (MP)}

In Middle-Point thresholding we consider the set of all power values from appliance $\ell$ in the training set $\{P_j^{(\ell)}\}_{j=1}^{n_{tot}}$. We apply a clustering algorithm to split this set into two clusters and consider the centroid of each cluster. Typically, the $k$-means clustering algorithm can be applied for this purpose,~\cite{macqueen1967kmeans}. The two centroids for each class, after applying $k$-means, are denoted by $m_0^{(\ell)}$ (for the OFF state) and $m_1^{(\ell)}$ (for the ON state).
In Middle-Point thresholding (MP), the threshold for appliance $\ell$ is fixed at the middle point between these values
\begin{equation}
    %\lambda^{()} = \frac{m_{0}^{(j)} + m_{1}^{(j)}}{2}
    \lambda^{(\ell)} = \frac{m_0^{(\ell)} + m_1^{(\ell)}}{2}.
    \label{eq:mp_threshold}
\end{equation}

\subsection{Variance-Sensitive Thresholding (VS)}

Variance-Sensitive (VS) thresholding was recently proposed as a finer version of MP thresholding by Desai et al.,\cite{desai2019multistate}. It also employs $k$-means clustering to find the centroids for each class, but the determination of the threshold now takes into account not only the mean, but also the standard deviation $\sigma^{(\ell)}_{k}$ for the points in each cluster, according to the following formula

\begin{align}
\begin{split}
   % d^{(j)} &= \frac{\sigma_{0}^{(j)}}{\sigma_{0}^{(j)} + \sigma_{1}^{(j)}} \\
    %\lambda^{(j)} &= (1 - d^{(j)}) \cdot m_{0}^{(j)} + d^{(j)} \cdot m_{1}^{(j)}
    d &= \frac{\sigma^{(\ell)}_0}{\sigma^{(\ell)}_0 + \sigma^{(\ell)}_{1}} \\
    \lambda^{(\ell)} &= (1-d) m_0^{(\ell)} + d m_1^{(\ell)}
\end{split}
\label{eq:vs_threshold}
\end{align}

\begin{figure}[ht]
\centering
  \includegraphics[trim={0 10.7cm 0 0}, clip, width=.9\linewidth]{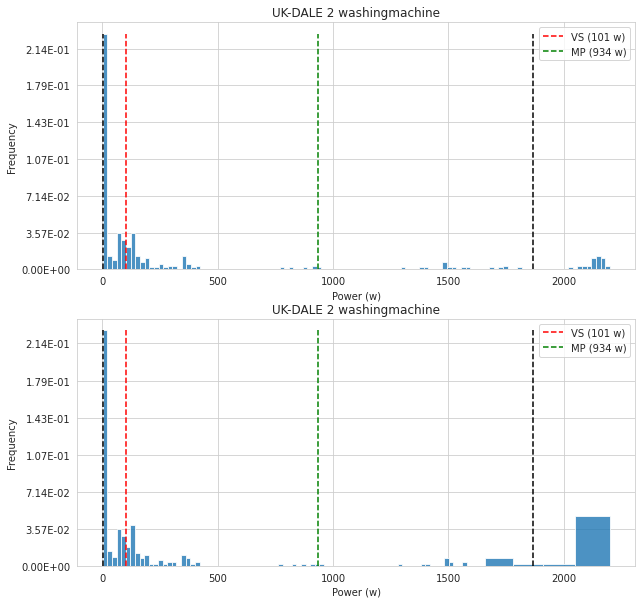}
  \caption{Distribution of a washing machine power load through all the monitoring time. The records pertain to house 2 of UK-DALE dataset. The graph was cropped vertically as 0 W consumption is more than $17\%$ of the total number of records. Black dashed lines mark the centroids found by $k$-means clustering (1 watts and 1866 watts), while color dashed lines are the thresholds fixed by different methods.}
  \label{fig:mp_vs}
\end{figure}

The motivation is that, if $\sigma_{1} > \sigma_{0}$, then the threshold should move towards $m_0$ in order not to misclassify the points in class $1$ that are further away from the centroid $m_1$. As a matter of fact, points in the OFF cluster usually have less variance, so the VS approach often sets its threshold lower than the MP approach. A comparison of both thresholding methods on a specific set of power measurements can be seen in Figure \ref{fig:mp_vs}. Also note that MP is a particular case of VS when $\sigma_{0} = \sigma_{1}$.

\subsection{Activation-Time Thresholding}

The last two methods only use data from the distribution of power measurements in order to fix the threshold for an appliance. It often happens that due to noise in the smart meters or devices, some measurements during short time intervals are either absent while the device is operating, or produce abnormal peaks during the OFF state. For this reason, to ensure a smoother behaviour, Kelly and Knottenbelt~\cite{kelly2015neural} set both a power threshold and a time threshold. The power threshold could be fixed by MP or VS or fixed externally by hand as done in ~\cite{kelly2015neural}. The time threshold $(\mu^{(\ell)}_0,\mu^{(\ell)}_1)$ specifies the minimum length of time that device $\ell$ must be in a given state, e.g. if a sequence of power measurements are below $\lambda^{(\ell)}$ for a time $t< \mu^{(\ell)}_0$, then that sequence is considered to be in the previous state (ON in this binary case).

In~\cite{kelly2015neural}, both power and time thresholds are chosen empirically, after analysing the appliance behaviour. Table \ref{tab:activation_time} shows the values of the thresholds relevant to our work. The threshold $\lambda$  is chosen usually at lower values, as the time threshold already filters noisy records. It would be desirable to turn this thresholding method into a fully automated data driven algorithm, in order to remove all subjective inputs.

\begin{table}[ht]
    \centering
    \begin{tabular}{|c|c|c|c|}
    \hline
        Threshold & Dishwasher & Fridge & Washing machine \\ \hline
        $\lambda$ (W) & 10 & 50 & 20\\ \hline
        $\mu_0$ (s) & 30 & 1 & 3\\ \hline
        $\mu_1$ (s) & 30 & 1 & 30 \\ \hline
    \end{tabular}
    \caption{Activation time (AT) threshold values used in this work, taken from~\cite{kelly2015neural}.}
    \label{tab:activation_time}
\end{table}

\begin{figure}[ht]
\centering
\begin{subfigure}{\textwidth}
  \centering
  \includegraphics[scale=0.5]{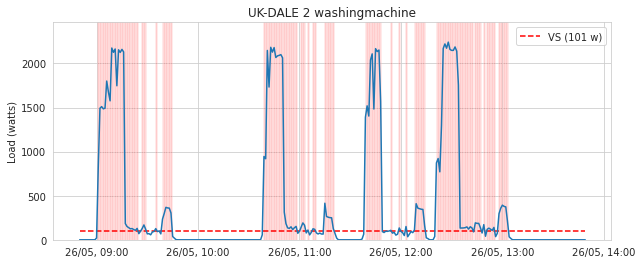}
  \label{fig:wm_vs}
\end{subfigure}
\begin{subfigure}{\textwidth}
  \centering
  \includegraphics[scale=0.5]{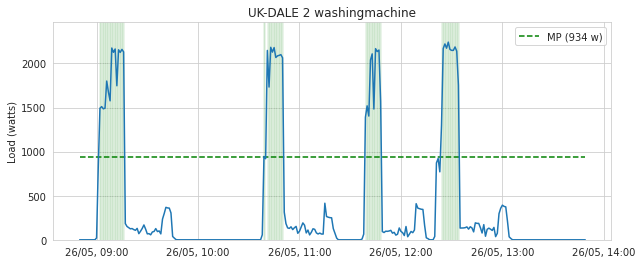}
  \label{fig:wm_mp}
\end{subfigure}
\begin{subfigure}{\textwidth}
  \centering
  \includegraphics[scale=0.5]{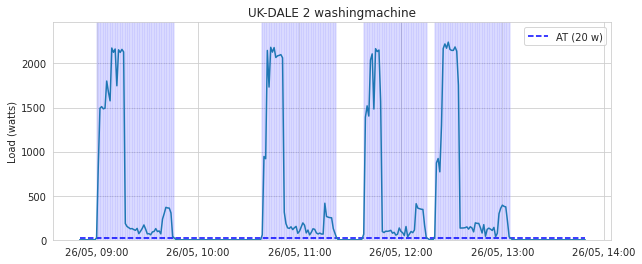}
  \label{fig:wm_at}
\end{subfigure}
\caption{Sample from the washing machine power load sequence, depicting how different thresholding methods classify each instance as ON or OFF.}
\label{fig:threshold_status}
\end{figure}

Figure \ref{fig:threshold_status} compares the three thresholding methods. Each graph shows the three values of $\lambda^{(\ell)}$ for a given device, together with the result of applying each thresholding method to the same input series. Observe that the same power data gives rise to rather different series for the ON/OFF status depending on the choice of thresholding method. We thus see that there are multiple ways to define a classification problem given the input signal.

A comparison and a discussion of each thresholding method after training state-of-the-art NILM for regression and classification problems will be performed in Section~\ref{sec:results}.

%%%%%%%%%%%%%%%%%%%%%%%%%%%%%%%%%%%%%%%%%%
\section{Neural Networks}\label{sec:models}
%%%%%%%%%%%%%%%%%%%%%%%%%%%%%%%%%%%%%%%%%%%%%

Almost all state of the art models propagate their inputs through one or more convolutional layers~\cite{depenha2017convolutional, yang2020multilabel}. This is done to ensure that the models are translation invariant. As NILM is related to time series, many studies also add recurrent layers (e.g. LSTM or GRU) to their networks~\cite{Kim2017NonintrusiveLM, kasemili2019bilstm}. These layers tend to get very good results on sequence-related problems. In this work, we will try out two different models: one that relies purely on convolutions, and other that also applies recurrent layers after the convolutions.

It is important to stress that both of these neural networks can be applied to train a classification model to predict device status or a regression model to predict device power load, the only difference in their architecture lies in the last layer (see Figure~\ref{fig:biGRU}), where an additional softmax layer needs to be added for the classification problem.
\subsection{Convolutional Network}

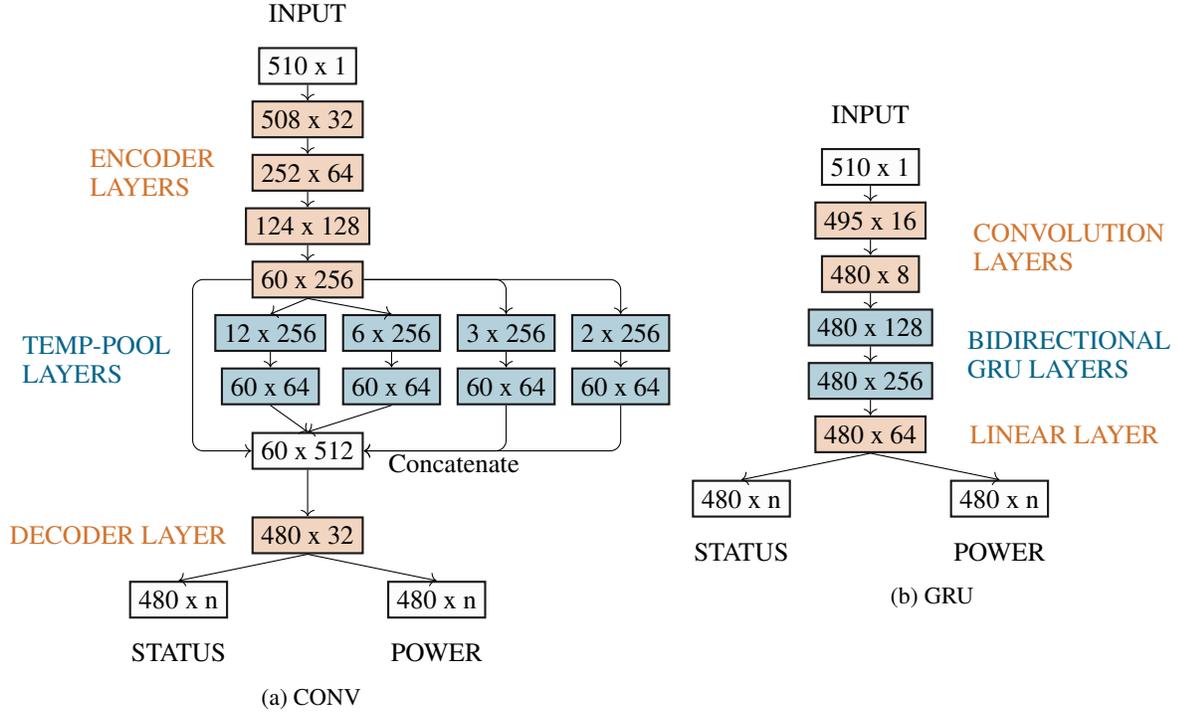
\begin{figure}[ht]
\centering
\begin{subfigure}{.5\linewidth}
\centering
%\resizebox{\linewidth}{!}{%
\begin{tikzpicture}[
node distance=0.025\textwidth,
squarednode/.style={rectangle, draw=black!90, fill=black!1, thick, minimum size=0.04\textwidth}
]

\node[squarednode](input){510 x 1};
\node[]()[above=of input]{INPUT};

% Encoder

\node[squarednode](conv1)[below=of input, fill=UCAOrange!30]{508 x 32};
%\node[]()[left=of conv1]{Conv};
\draw[->](input.south)--(conv1.north);

\node[squarednode](conv2)[below=of conv1, fill=UCAOrange!30]{252 x 64};
%\node[]()[left=of conv2]{Convolution};
\draw[->](conv1.south)--(conv2.north);

\node[squarednode](conv3)[below=of conv2, fill=UCAOrange!30]{124 x 128};
%\node[]()[left=of conv3]{Convolution};
\draw[->](conv2.south)--(conv3.north);

\node[squarednode](conv4)[below=of conv3, fill=UCAOrange!30]{60 x 256};
%\node[]()[left=of conv4]{Convolution};
\draw[->](conv3.south)--(conv4.north);

\node[]() at ($(conv1)!0.5!(conv3)$)[xshift=-.25\linewidth, align=left, text=UCAOrange]{ENCODER\\LAYERS};

% Decoder

\node[squarednode](tpool1)[below=of conv4, xshift=-.06\textwidth, fill=UCABlue!30] {12 x 256};
\draw[->](conv4.south)--(tpool1.north);
\node[squarednode](tpool2)[right=of tpool1, fill=UCABlue!30]{6 x 256};
\draw[->](conv4.south)--(tpool2.north);
\node[squarednode](tpool3)[right=of tpool2, fill=UCABlue!30]{3 x 256}; 
\draw[rounded corners, ->] (conv4.east) -| (tpool3.north);
\node[squarednode](tpool4)[right=of tpool3, fill=UCABlue!30]{2 x 256};
\draw[rounded corners, ->] (conv4.east) -| (tpool4.north);

\node[squarednode](ups1)[below=of tpool1, fill=UCABlue!30] {60 x 64};
\draw[->](tpool1.south)--(ups1.north);
\node[squarednode](ups2)[below=of tpool2, fill=UCABlue!30] {60 x 64};
\draw[->](tpool2.south)--(ups2.north);
\node[squarednode](ups3)[below=of tpool3, fill=UCABlue!30] {60 x 64};
\draw[->](tpool3.south)--(ups3.north);
\node[squarednode](ups4)[below=of tpool4, fill=UCABlue!30] {60 x 64};
\draw[->](tpool4.south)--(ups4.north);

\node[]()  at ($(tpool1)!0.5!(ups1)$)[xshift=-.28\textwidth, align=left, text=UCABlue]{TEMP-POOL\\LAYERS};

% Decoder

\node[squarednode](concat)[below=of conv4, yshift=-0.19\textwidth] {60 x 512};
\node[]()[right=of concat, yshift=-.02\textwidth]{Concatenate};

\draw[->](ups1.south)--(concat.north);
\draw[->](ups2.south)--(concat.north);
\draw[rounded corners, ->] (ups3.south) |- (concat.east);
\draw[rounded corners, ->] (ups4.south) |- (concat.east);

\node[](fake1)[left=of tpool1]{};
\node[](fake2)[left=of ups1]{};
\draw[rounded corners, ->] 
(conv4.west) -| ($ (fake2)!.5!(fake1) $) |- (concat.west);

\node[squarednode](decoder)[below=of concat, yshift=-.05\textwidth, fill=UCAOrange!30]{480 x 32};
\node[]()[left=of decoder, align=left, text=UCAOrange]{DECODER LAYER};
\draw[->](concat.south)--(decoder.north);

% Output

\node[squarednode](activation)[below left=of decoder, yshift=-.025\textwidth, xshift=-.02\textwidth] {480 x n};
\node[]()[below=of activation]{STATUS};
\draw[->](decoder.south)--(activation.north);

\node[squarednode](regression)[below right=of decoder, yshift=-.025\textwidth, xshift=.02\textwidth] {480 x n};
\node[]()[below=of regression]{POWER};
\draw[->](decoder.south)--(regression.north);

\end{tikzpicture}

%}%

\caption{CONV}
\label{fig:tpnilm}
\end{subfigure}%
\begin{subfigure}{.5\linewidth}
\centering
%\resizebox{\linewidth}{!}{%
\begin{tikzpicture}[
node distance=0.025\textwidth,
squarednode/.style={rectangle, draw=black!90, fill=black!1, thick, minimum size=0.04\textwidth}
]

\node[squarednode](input){510 x 1};
\node[]()[above=of input]{INPUT};

\node[squarednode](conv1)[below=of input, fill=UCAOrange!30]{495 x 16};
\draw[->](input.south)--(conv1.north);

\node[squarednode](conv2)[below=of conv1, fill=UCAOrange!30]{480 x 8};
\draw[->](conv1.south)--(conv2.north);

\node[]() at ($(conv1)!0.5!(conv2)$)[xshift=.32\textwidth, align=left, text=UCAOrange]{CONVOLUTION\\LAYERS};

\node[squarednode](gru1)[below=of conv2, fill=UCABlue!30]{480 x 128};
\draw[->](conv2.south)--(gru1.north);

\node[squarednode](gru2)[below=of gru1, fill=UCABlue!30]{480 x 256};
\draw[->](gru1.south)--(gru2.north);

\node[]() at ($(gru1)!0.5!(gru2)$)[xshift=.32\textwidth, align=left, text=UCABlue]{BIDIRECTIONAL\\GRU LAYERS};

\node[squarednode](dense)[below=of gru2, fill=UCAOrange!30]{480 x 64};
\node[]()[right=of dense,xshift=0.03\linewidth, text=UCAOrange]{LINEAR LAYER};
\draw[->](gru2.south)--(dense.north);

% Output

\node[squarednode](activation)[below left=of dense, yshift=-.025\textwidth, xshift=-.02\textwidth] {480 x n};
\node[]()[below=of activation]{STATUS};
\draw[->](dense.south)--(activation.north);

\node[squarednode](regression)[below right=of dense, yshift=-.025\textwidth, xshift=.02\textwidth] {480 x n};
\node[]()[below=of regression]{POWER};
\draw[->](dense.south)--(regression.north);

\end{tikzpicture}
%}%
\caption{GRU}
\label{fig:biGRU}
\end{subfigure}
\caption{Architecture of each model. There are two possible outputs, the model trains differently depending of which output we chose.}
\label{fig:models}
\end{figure}

Our first model relies solely on  convolutional layers, inspired on the architecture from the work of Luca Massidda et al.~\cite{massidda2020tpnilm}. The general scheme follows the classic approach to the semantic segmentation of images. See Figure \ref{fig:tpnilm} to better understand the following model explanation.

The CONV model receives  as input a vector with size $L_{\textrm{in}} = 510$ which represents the household aggregated power over an $8\frac{1}
{2}$ hour interval. The vector is propagated through an encoder, characterised by alternating convolution and pooling modules. Each encoder layer begins with a convolution of kernel size 3 and no padding, then applies batch normalization and ReLU activation, and ends with a max pooling of kernel size 2 and stride 2. Only the last layer omits the max pooling step. Encoder layers increase the space of the features of the signal at the cost of decreasing the temporal resolution.

After that, the \textit{Temporal Pooling} module aggregates the features at different resolutions, which is reminiscent of inception networks~\cite{szegedy2014inception}. Four different average poolings are applied, with kernel sizes  5, 10, 20 and 30; having the same stride as kernel size. Each of those layers then propagate their values through convolution layers of kernel size 1 and no padding, followed by a batch normalization and ReLU activation. All of their outputs are then concatenated.

Finally, the decoder module applies one convolution of kernel size 8 and stride 8, followed by batch normalization. It then bifurcates into two different outputs: the appliance status and the appliance power. Both outputs are computed by propagating the network values through one last convolution layer of kernel size 1 and padding 1. In the case of status, we apply the softmax function. Both status and power load output vectors have the same sampling frequency as the input aggregate load, but have a shorter length $L_{\textrm{out}} = 480$ as explained in Section~\ref{sec:methodology}. 

\subsection{Bidirectional GRU Network}

Some authors tend to connect convolutional and recurrent layers to extract  temporal correlations out of the input sequence~\cite{kyrstalakos2018sliding, kelly2015neural}. This second model follows a prototypical GRU scheme, depicted in Figure \ref{fig:biGRU}.
The input and output layers are the same as in the previous model. For the processing units, the GRU model propagates the input vector through two convolutional layers with kernel size 20, padding 2 and stride 1, before applying the recurrent (bidirectional GRU) layers. 

One can see that the model architecture is rather lightweight. However, GRU takes longer to train than CONV: adapting the GRU weights requires a lot of computation, compared to updating the weights of convolutional layers.

%%%%%%%%%%%%%%%%%%%%%%%%%%
\section{Methodology}\label{sec:methodology}
%%%%%%%%%%%%%%%%%%%%%%%%%%%%%%%%%%%%%%%

\subsection{Preprocessing}

In order to make our results reproducible and easy to compare with other works, we restrict to the  UK-DALE dataset~\cite{kelly2015ukdale}, which is a standard benchmark for NILM. 

\begin{table}[ht]
\centering
\begin{tabular}{|l|l|}
\hline
Date of release & 2014 \\ \hline
Location & United Kingdom \\ \hline
Number of households & 5 \\ \hline
Meter units & Watts \\ \hline                                             
Sampling frequency & 6 seconds \\ \hline
\end{tabular}
\caption{UK-DALE dataset features.}
\label{tab:ukdale}
\end{table}

\begin{table}[ht]
\centering
\begin{tabular}{|l|c|c|c|c|c|c|}
\hline
\textbf{Building} & 1 & 2 & 3 & 4 & 5 \\ \hline
\textbf{Total time (days)} & $>$365 & 235 & 39 & 206 & 137  \\ \hline
\textbf{Appliances} & 53 & 18 & 4 & 11 & 24 \\ \hline
\end{tabular}
\caption{UK-DALE dataset.}
\label{table:ukdale_houses}
\end{table}

Only houses 1, 2 and 5 have been used for this work. Our target appliances which are found in the three buildings are: fridge, dishwasher and washing machine. This choice of houses and appliances is common in other works~\cite{massidda2020tpnilm, kelly2015neural, shin2019subtask}, as they seek to monitor appliances with distinguishable load absorption patterns and relevant contribution to the total power consumption.

Every power load series was downsampled from 6 seconds to 1 minute-frequency. After this downsampling, every input sequence comprises $8\frac{1}{2}$ hours of time, which amount to $L_{\rm in} = 510$  records. Since the models use convolutions with no padding, the first and last records of each series are dropped in the output, thus leading to an output sequence having $L_{\rm out} = 480$ records, i.e. 8 hours. We have divided the original time series into input sequences with an overlap of 30 records between consecutive input sequences, so that the output sequence are continuous in time and have no gaps. Aggregate power load is normalized, dividing the load by a reference power value of 2000 W for numerical stability. Each input series is further normalized by subtracting its mean.  Thus, we can define the following input and output series for the regression problem
\begin{flalign}
  \text{Regression:}  \qquad  &\boldsymbol{x}_j = \frac{\boldsymbol{P}_j - \Bar{P_j}}{2000} , \qquad
    %\boldsymbol{y^{(j)}} = \frac{\boldsymbol{p^{(j)}}}{2000}
    \boldsymbol{y}^{(\ell)}_j = \frac{\boldsymbol{P}_j^{(\ell)}}{2000},
\end{flalign}
where $\Bar{P_j}=\displaystyle\frac{1}{L_{\rm in}}\sum_{i=1}^{L_{\rm in}} P_{j,i}$.

To define the classification problem, the target $\boldsymbol{y}^{(\ell)}_j$ is the device status which is computed from ${\bf P}^{(\ell)}_j$ using the thresholding methods described in Section~\ref{sec:threshold}. More specifically, we have
\begin{flalign}
  \text{Classification:}  \qquad  &\boldsymbol{x}_j = \frac{\boldsymbol{P}_j - \Bar{P_j}}{2000},  \qquad
    %\boldsymbol{y^{(j)}} = \frac{\boldsymbol{p^{(j)}}}{2000}
    \boldsymbol{y}^{(\ell)}_j = \boldsymbol{s}^{(\ell)}_j ,
\end{flalign}
where $\boldsymbol{s}^{(\ell)}_j $ is a series of binary values defined by \eqref{eq:status}.

The training set for each problem is built by adding the first $80\%$ sequences from each of the three buildings, which amounts to 1941 sequences (describing a total time of 687 days of measurements). The validation set is built by using the subsequent $10\%$ records from house UK-DALE 1, for a total of 183 sequences (65 days), while the test set is composed of the last $10\%$ sequences of the same building, having the same size as the validation set.

\begin{table}[ht]
    \centering
    \begin{tabular}{|c|c|c|c|}
    \hline
    & Train & Validation & Test  \\ \hline
    Number of points ($N$) & 1941 & 183 & 183 \\ \hline
    Total time (days) & 687 & 65 & 65 \\ \hline
    \end{tabular}
    \caption{Training, validation and test sets. Each point is a pair ($\boldsymbol{x}_j$, $\boldsymbol{y}^{(\ell)}_j)$ of time series, where $\boldsymbol{x}_j$ has $L_{\rm in} = 510$ records, and  $\boldsymbol{y}^{(\ell)}_j$ has $L_{\rm out} = 480$ records, both of them at 1 min. intervals.}
    \label{tab:data_split}
\end{table}

In order to judge whether we are dealing with a balanced classification problem, it will be useful to report how often a given device has been ON during the training and test sets, depending on which thresholding method has been applied. The results can be seen in Table~\ref{tab:frequency}.

\begin{table}[ht]
    \centering
    \begin{tabular}{|c|c|c|c|c|}
    \hline
        \textbf{Threshold} & \textbf{Set} & \textbf{Dishwasher} & \textbf{Fridge} & \textbf{Washing Machine} \\ \hline
        \multirow{2}{*}{MP} & Train & 0.77 & 42 & 0.97 \\ \cline{2-5}
         & Test & 0.62 & 46 & 1.70 \\ \hline
         \multirow{2}{*}{VS} & Train & 0.83 & 44 & 1.80 \\ \cline{2-5}
         & Test & 0.67 & 47 & 2.50 \\ \hline
         \multirow{2}{*}{AT} & Train & 2.30 & 42 & 4.60 \\ \cline{2-5}
         & Test & 2.40 & 45 & 7.30 \\ \hline
    \end{tabular}
    \caption{Fraction of activation time (in \%) for each device over the train and test sets.}
    \label{tab:frequency}
\end{table}

It is important to stress a number of things from the observation of this table. First, the fraction of ON states is clearly dependent on the thresholding method, as it was already clear from Figure~\ref{fig:threshold_status}. Next, dishwasher and washing machine are only sparsely activated, while the fridge is considered to be ON roughly half of the time. In these two cases we are dealing with an imbalanced class problem, which should be taken into account when defining and interpreting the appropriate metrics. Finally, in all cases but specially for the washing machine, the prevalence of the positive class differs greatly from the training to the test set. This of course could happen because these periods have been chosen consecutive in time (in accordance with other works). If the train-validation-test split was done randomly over the 2307 series records, we would observe a similar class distribution across sets. Splitting training and test sets for a time series in machine learning problems always involves a delicate choice: whether to split records randomly (which ensures homogeneous distribution) or chronologically (which is closer to the real operating conditions).

\subsection{Training}
Each of the models described in Section~\ref{sec:models} was trained for 300 epochs. Training data was fed to the model in batches of 32 sequences, shuffled randomly. 

The loss function for the regression problem is the mean square error or L2 metric, given by

\begin{equation}
    \mathcal{L}^{(\ell)}_{\rm reg}=\frac{1}{\Nt}\sum_{j=1}^{\Nt}\frac{1}{L_{\rm out} }\sum^{L_{\rm out} }_{i=1} \left( y^{(\ell)}_{j,i}-\hat y^{(\ell)}_{j,i} \right)^2.\qquad 
    \label{eq:regloss}
\end{equation}
The standard choice of loss function for the classification problem is binary cross entropy:
\begin{equation}
    \mathcal{L}^{(\ell)}_{\rm class}=\frac{1}{\Nt}\sum_{j=1}^{\Nt}\frac{1}{L_{\rm out} }\sum^{L_{\rm out} }_{i=1} \left( y^{(\ell)}_{j,i}\cdot \log\hat y^{(\ell)}_{j,i} + (1- y^{(\ell)}_{j,i})\cdot \log (1- \hat y^{(\ell)}_{j,i})   \right),\qquad 
    \label{eq:classloss}
\end{equation}
where $\hat{\boldsymbol{y}}^{(\ell)}_j$  is the probability that device $\ell$ is ON at each time step.

During the 300 training epochs, we keep the model that achieves the minimum loss over the validation set, using an Adam optimizer for weights update, with a starting learning rate of $1E-4$.

Both data preprocessing and neural network training were performed on Python. Specifically, the models were written on Pytorch and trained in a GPU NVidia GeForce GTX 1080 with 8 GB of VRAM, NVIDIA-SMI 440.95.01 and CUDA v10.2. The code for this paper is available online\footnote{Open-source repository: \url{https://github.com/UCA-Datalab/better_nilm}} and the data comes from a public dataset, so all results reported in this paper are reproducible. Using the configuration stated in this section, CONV models took from 7 to 8 minutes to train 300 epochs, while GRU architectures took 16 minutes.

\subsection{Metrics}\label{sec:metrics}

Although metrics are clearly related to loss functions used for model training, the main difference is that the reported metrics are not required to be differentiable. 
When the output is a continuous variable (power load in our case), we use as the relevant metric the L1 error or MAE rather than RMSE, since the latter tends to give too much importance to large deviations.
\begin{equation}
 \text{MAE}^{(\ell)}=\frac{1}{\Nt}\sum_{j=1}^{\Nt}\frac{1}{L_{\rm out} }\sum^{L_{\rm out} }_{i=1} \left| y^{(\ell)}_{j,i}-\hat y^{(\ell)}_{j,i} \right|,\qquad 
\end{equation}

When the output variable is categorical, we can use the $F_1$-score to balance precision and recall in balanced problems where there is no preference to achieve a better classification of the ON or OFF classes. The $F_1$-score is the harmonic mean of precision and recall, or directly
\begin{equation}
    F_{1}^{(\ell)} = \frac{1}{N_{\rm{test}}}\sum_{j=1}^{N_{\rm{test}}}\frac{TP_j^{(\ell)}}{TP_j^{(\ell)} + \frac{1}{2}(FP_j^{(\ell)} + FN_j^{(\ell)})}
    \label{eq:f1}
\end{equation}
where the true positives (TP), false positives (FP), true negatives (TN) and false negatives (FN) in each series are given by
\begin{align}
    &TP^{(\ell)}_j = \sum^{L_{\rm out} }_{i=1} \hat{s}^{(\ell)}_{j,i} \cdot y^{(\ell)}_{j,i}, \nonumber&
    &FP^{(\ell)}_j = \sum^{L_{\rm out} }_{i=1} \hat{s}^{(\ell)}_{j,i} \cdot (1 - y^{(\ell)}_{j,i}), \nonumber \\
    &FN^{(\ell)}_j = \sum^{L_{\rm out} }_{i=1} (1 - \hat{s}^{(\ell)}_{j,i}) \cdot y^{(\ell)}_{j,i}, \nonumber&
    &TN^{(\ell)}_j = \sum^{L_{\rm out} }_{i=1} (1 - \hat{s}^{(\ell)}_{j,i}) \cdot (1 - y^{(\ell)}_{j,i}).
\end{align}
where in this case $y^{(\ell)}_{j,i}$ is the state of device $\ell$ at instant $i$, and $\hat s^{(\ell)}_{j,i}$ is the predicted state as defined by 
\begin{equation}
    \hat{s}^{(\ell)}_{j,i} = I \left(\hat{y}^{(\ell)}_{j,i} - 0.5\right).
\end{equation}

For imbalanced problems, such as NILM for the dishwasher and washing machine (see Table~\ref{tab:frequency}), the $F_1$-score might not be the best metric to report. Rather, since those devices are ON roughly 1\% of the time, it is better to consider a suitable metric for imbalanced classification problems, like the area under the ROC curve.~\cite{fawcett2006introduction}.

%%%%%%%%%%%%%%%%%%%%%%%%%%%%%%%%%%%
\section{Results}\label{sec:results}
%%%%%%%%%%%%%%%%%%%%%%%%%%%%%%%%%%%5

\subsection{Regression problem}
We begin our discussion of results by reporting the metrics obtained over the test set by the two models (CONV and GRU) in the regression problem for each of the three appliances $\ell=\{\text{dishwasher},\text{fridge},\text{washing machine}\}$.

\begin{table}[ht]
    \centering
    \begin{tabular}{|c|c|c|c|}
    \hline
        \textbf{Model} & \textbf{Dishwasher} & \textbf{Fridge} & \textbf{Washing Machine} \\ \hline
        CONV & 11.59 & 26.95 & 18.25 \\ \hline
        GRU & 8.07 & 28.68 & 15.00 \\ \hline
    \end{tabular}
    \caption{MAE scores (in watts) for regression models on each appliance}
    \label{tab:regression_mae}
\end{table}

The analysis of these metrics needs a word of caution: it is natural to expect that if a device has been sparsely activated during the test set, the predicted load will give an overall lower MAE than a similar device that has been used more often (see Table~\ref{tab:frequency}). For this reason, some authors have suggested other metrics such as the energy error~\cite{mayhorn2016load, batra2014nilmtk}. For a recent review of all metrics that have been used in NILM, see~\cite{pereira2018metrics}.

\begin{figure}[ht]
%\centering
\begin{subfigure}{.5\linewidth}
  \centering
  \includegraphics[trim=5 0 5 0, clip, width=1.1\linewidth]{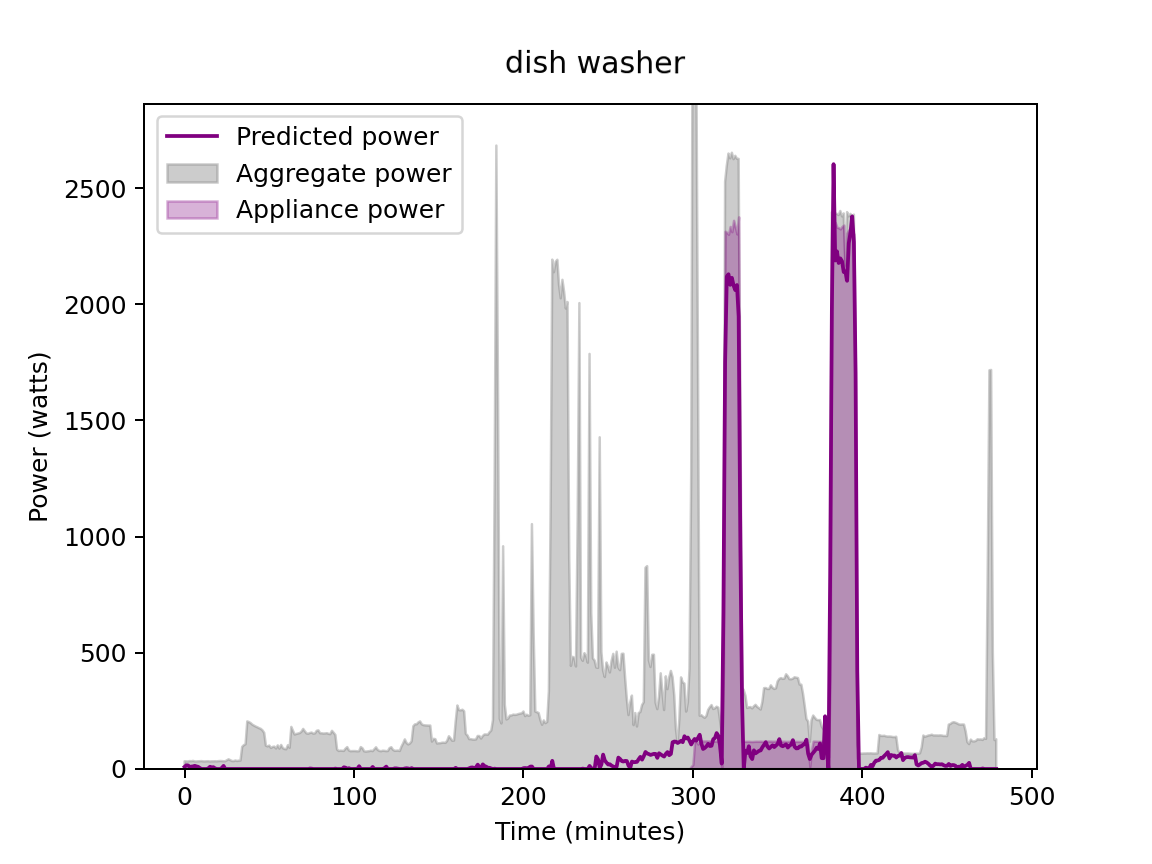}
\end{subfigure}
\begin{subfigure}{.5\linewidth}
  \centering
  \includegraphics[trim=5 0 5 0, clip, width=1.1\linewidth]{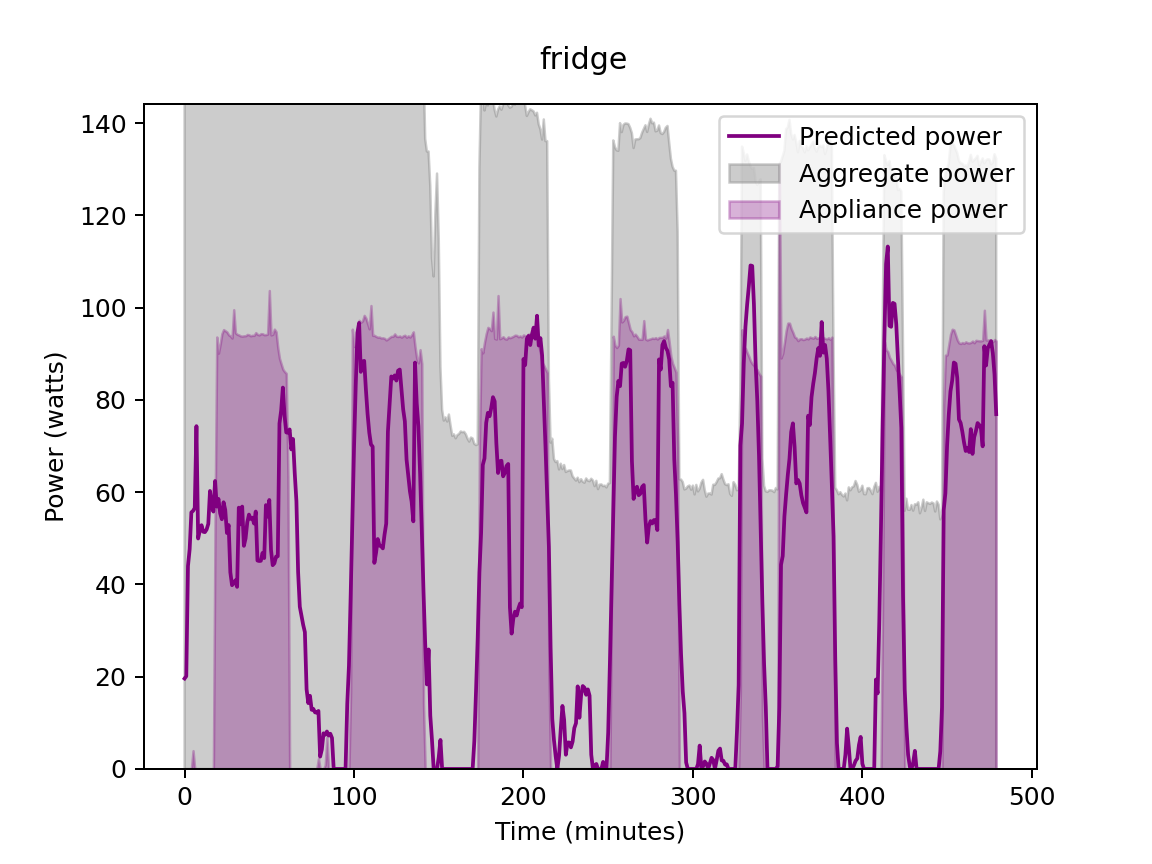}
\end{subfigure}
\begin{subfigure}{\linewidth}
  \centering
  \includegraphics[width=.55\linewidth]{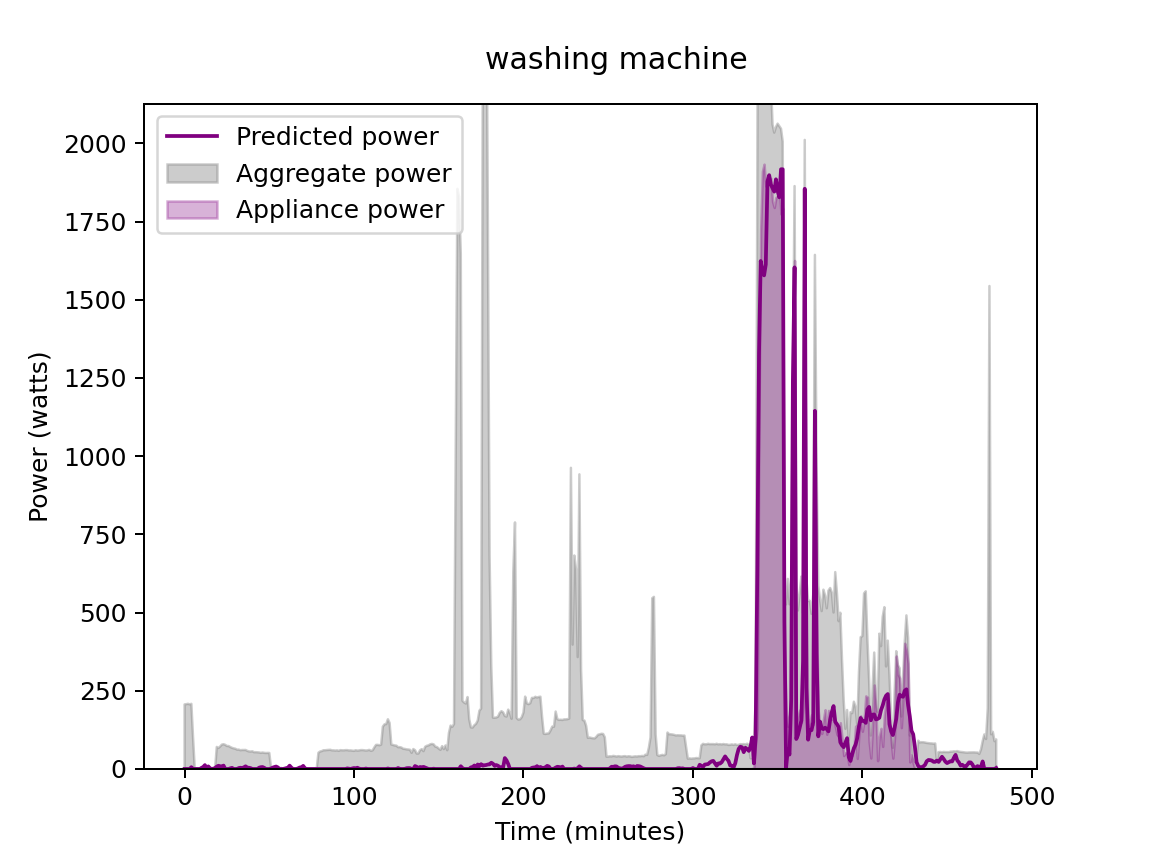}
\end{subfigure}
\caption{CONV regression: aggregated power load (input signal) and real and predicted loads for each device.}
\label{fig:regression}
\end{figure}

To understand the complexity of disaggregating the power signal in NILM, we have plotted a time series from the test set in Figure~\ref{fig:regression}. In the plots we show the input signal $\mathbf{P}_j$ (aggregated power load), together with the real power of each device $\mathbf{P}^{(\ell)}_j$ and the predicted power obtained from the CONV network $\hat{\mathbf{P}}^{(\ell)}_j$.

In the first graph of Figure~\ref{fig:regression} we see that the model has identified correctly the two main power peaks of 2200 W corresponding to the dish washer, but has properly ignored other similar large peaks occurring earlier in the series. In the second graph, we observe that the fridge power prediction is often masked by the presence of other devices, having a mean value of just 100 W but a very periodic activation pattern. When the aggregate power load is small, the model is able to better resolve the signal coming from the fridge. The washing machine has a more complex consumption pattern during its activation period, but the model has also been able to identify correctly the activation peaks while ignoring other similar peaks of the same magnitude occurring earlier in the series that do not correspond to washing machine operation.

\subsection{Classification problem}

As we have mentioned above, the classification problem is not uniquely defined since the raw data do not include the real intervals where each device was ON/OFF, but only its consumption. Thus, we have three different classification problems depending on the choice of thresholding methods described in Section~\ref{sec:threshold}.
For each possible value of $(\text{model}, \text{thresholding}, \text{device})$ we report the $F_1$-score \eqref{eq:f1} over the test set in Table~\ref{tab:classification_f1}.

\begin{table}[ht]
    \centering
    \begin{tabular}{|c|c|c|c|c|}
    \hline
        \textbf{Threshold} & \textbf{Model} & \textbf{Dishwasher} & \textbf{Fridge} & \textbf{Washing Machine} \\ \hline
        \multirow{2}{*}{MP} & CONV & 0.93 & 0.87 & 0.93 \\ \cline{2-5}
         & GRU & 0.84 & 0.87 & 0.87 \\ \hline
         \multirow{2}{*}{VS} & CONV & 0.93 & 0.87 & 0.88 \\ \cline{2-5}
         & GRU & 0.84 & 0.87 & 0.82 \\ \hline
         \multirow{2}{*}{AT} & CONV & 0.91 & 0.86 & 0.97 \\ \cline{2-5}
         & GRU & 0.90 & 0.86 & 0.96 \\ \hline
    \end{tabular}
    \caption{$F_1$-scores for classification models on each appliance and threshold}
    \label{tab:classification_f1}
\end{table}

\begin{figure}[ht]
%\centering
\begin{subfigure}{.5\linewidth}
  \centering
  \includegraphics[trim=5 0 5 0, clip, width=\linewidth]{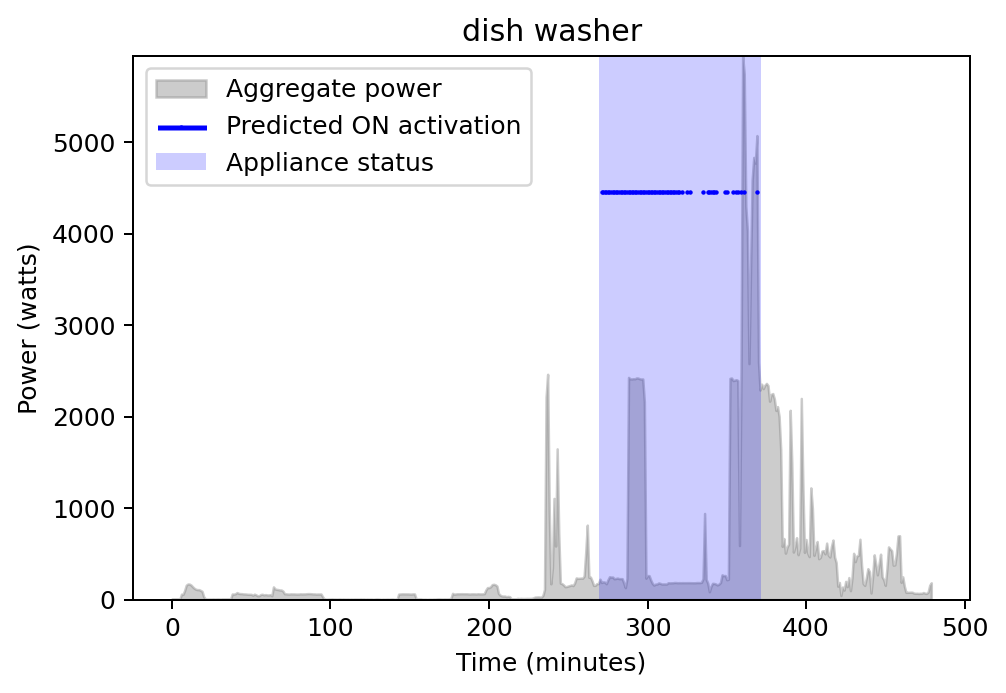}
\end{subfigure}
\begin{subfigure}{.5\linewidth}
  \centering
  \includegraphics[trim=5 0 5 0, clip, width=\linewidth]{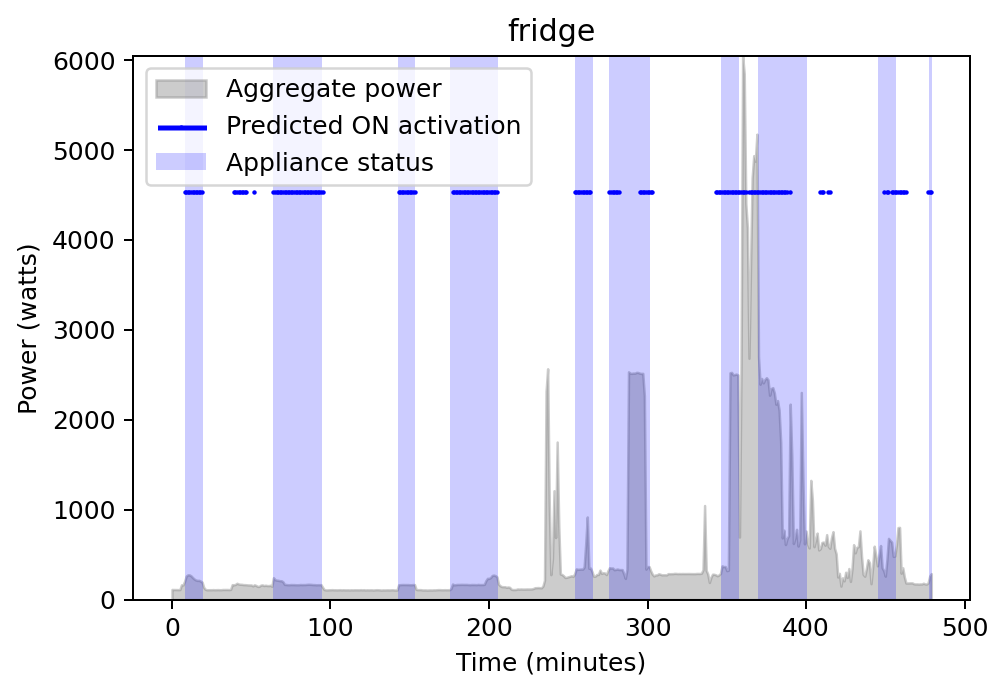}
\end{subfigure}
\begin{subfigure}{\linewidth}
  \centering
  \includegraphics[width=.5\linewidth]{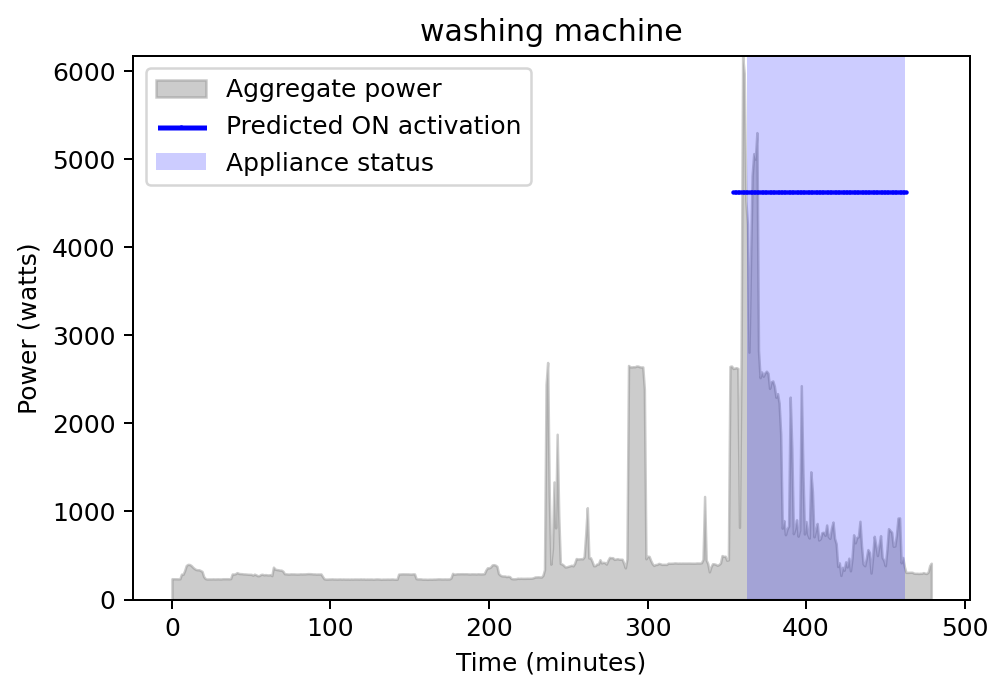}
\end{subfigure}
\caption{CONV model applying AT threshold: aggregated power load (input signal), real device status  (output signal) and predicted status.}
\label{fig:classification}
\end{figure}

The next to last line shows that our results are in good agreement with those reported by the authors of~\cite{massidda2020tpnilm}. Also, we can observe that CONV has a slightly better $F_1$-score that GRU for the classification problems in all three devices and thresholding methods, although both of them show a very good performance (see Figure~\ref{fig:classification}).
In general, the classification problem for the fridge is harder, for reasons that have been already mentioned above.
We raised the idea of using AUC over $F_1$-score in Section~\ref{sec:metrics}. We computed AUC and realized it was almost identical to $F_1$-score, so we will keep using the latter as it is more common in the literature.

It is also instructive to represent the input signal $\mathbf{P}_j$, together with the real output signal $\mathbf{P}^{(\ell)}_j$ describing the status of device $\ell$ and the predicted status $\mathbf{s}_j^{(\ell)}$, to grasp the nature of NILM problem for classification. We have plotted  in Figure~\ref{fig:classification}  the output of CONV on a given series of records from the test set where the three devices have been activated (sometimes simultaneously). Observe that the model is able to discriminate, with a very good precision, the periods in which each of the devices are activated, just from the observation of the aggregated power signal.

\subsection{Reconstructing the power signal}\label{sec:recon}

A very natural question to address is which of the three proposed thresholding methods should be preferred. One would naively think that the one leading to a better $F_1$-score would be the best choice, if this was only based on prediction performance. However, placing a trivial threshold of zero would yield state ON for all time intervals at the training and test sets, and any decent ML method would immediately learn this, thus reaching a perfect $F_1$-score but having no useful interpretation at all. Thus, we need to balance predictive performance with a way of judging which method is more meaningful. In the absence of any other external information on when each device can be considered to be ON or OFF, we need to find an objective quantitative argument to tackle this question.

For this purpose, we propose to reconstruct the power signal from each device and compare the reconstructed signal with the original power load of the device. We compute the average power $\bar{P}_{ON}$, (resp. $\bar{P}_{OFF}$) for device $\ell$ during the periods that are considered to be ON, (resp. OFF) after applying the thresholding method, and reconstruct the power series with these binary values. 

More specifically, we have
\begin{eqnarray}
\bar{P}_{ON}^{(\ell)}&=& \frac{1}{\Nt}\sum_{j=1}^{\Nt}\frac{1}{L_{\rm out} }\sum_{i=1}^{L_{\rm out} } s_{j,i}^{(\ell)} P_{j,i},\\
\bar{P}_{OFF}^{(\ell)}&=& \frac{1}{\Nt}\sum_{j=1}^{\Nt}\frac{1}{L_{\rm out} }\sum_{i=1}^{L_{\rm out} } \left(1-s_{j,i}^{(\ell)}\right) P_{j,i},
\end{eqnarray}
and reconstruct a binary power load for device $\ell$ as
\begin{equation}
\textbf{BP}_j^{(\ell)}=\bar{P}_{ON}^{(\ell)}\,\mathbf{s}_j^{(\ell)}+ \bar{P}_{OFF}^{(\ell)}\, (1-\mathbf{s}_j^{(\ell)})
\end{equation}

\begin{figure}[ht]
    \centering
    \includegraphics[width=.9\textwidth]{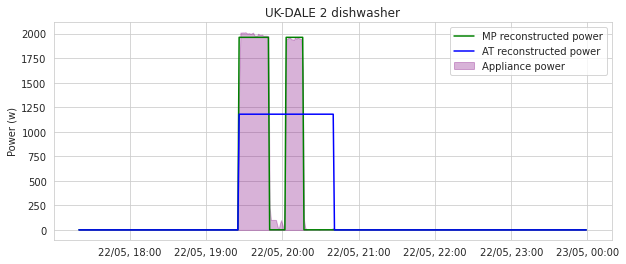}
    \caption{Device power load reconstruction from two different thresholding methods.}
    \label{fig:reconstruction}
\end{figure}

The reconstructed power series can be seen, together with the original series in Figure~\ref{fig:reconstruction} for two of the thresholding methods, corresponding to the same data as Figure~\ref{fig:threshold_status}.

We can compute the MAE between the original   $\mathbf{P}^{(\ell)}$ and reconstructed series $\textbf{BP}^{(\ell)}$ averages over the training set, which we call the {\it intrinsic error}, since it is prior to any prediction method. The results for the three devices and thresholding methods are shown in Table~\ref{tab:intrinsic}.
\begin{table}[ht]
    \centering
    \begin{tabular}{|c|c|c|c|}
    \hline
        Threshold & Dishwasher & Fridge & Washing machine \\ \hline
        MP & 3.48 & 4.67 & 3.96 \\ \hline
        VS & 4.39 & 4.71 & 6.60 \\ \hline
        AT & 26.37 & 4.66 & 7.42 \\ \hline
    \end{tabular}
    \caption{Intrinsic error: MAE between the original and reconstructed power series.}
    \label{tab:intrinsic}
\end{table}

From this comparison, we see that the Activation Time  (AT) thresholding is the one having the largest intrinsic error, while Middle Point (MP) thresholding offers the closest reconstructed power series. The fridge has similar intrinsic error for all three methods since the original series is very regular, being almost a binary series itself (see Figure~\ref{fig:regression}). 

As we mentioned above, it is not enough to look only at the classification metrics in Table~\ref{tab:classification_f1} to judge which is the best thresholding method for a NILM problem. For this reason, given the prediction output of the classification problem, we compute the reconstructed binary series and compare it with the original power series. MAE averaged over the test set are reported in Table~\ref{tab:regression_reconstructed} for the two models (GRU and CONV).

\begin{table}[ht]
    \centering
    \begin{tabular}{|c|c|c|c|c|}
    \hline
        \textbf{Threshold} & \textbf{Model} & \textbf{Dishwasher} & \textbf{Fridge} & \textbf{Washing Machine} \\ \hline
        \multirow{2}{*}{MP} & CONV & 9.09 & 21.45 & 39.45 \\ \cline{2-5}
         & GRU & 11.25 & 21.80 & 41.66 \\ \hline
         \multirow{2}{*}{VS} & CONV & 9.25 & 21.01 & 40.74 \\ \cline{2-5}
         & GRU & 11.59 & 22.19 & 41.85 \\ \hline
         \multirow{2}{*}{AT} & CONV & 19.75 & 21.26 & 57.23 \\ \cline{2-5}
         & GRU & 21.07 & 22.45 & 56.35 \\ \hline
    \end{tabular}
    \caption{MAE scores (in watts) for classification models after reconstructing the power load, on each appliance and threshold.}
    \label{tab:regression_reconstructed}
\end{table}
The first thing to note is that naturally the MAE errors are larger than the intrinsic errors, as they incorporate the errors in the classification. However, it is worth noting that for the dishwasher in AT thresholding this error is actually smaller than the intrinsic one. We find the explanation for that phenomenon in Figure~\ref{fig:reconstruction}: current AT parameters contain a large time threshold for the dishwasher (see Table~\ref{tab:activation_time}), which might lead to incorrect thresholding and large deviations in the reconstructed series. This suggests that the free parameters in AT thresholding should be optimized my minimizing the intrinsic error for each device over the training set. As shown in previous tables, CONV has a slightly better performance than GRU. It is also worth noting that the MAE for the washing machine has increased by a factor of 10 with respect to the intrinsic error, while in the other two devices, the factor is close to 4. The most likely explanation for this deviation is due to the train-test splitting: most of the error comes from activation periods, and these occur twice more often in the test than in the training set for the washing machine (see Table~\ref{tab:frequency}) while there is hardly no variation in the other two devices. This brings back the already mentioned remark on the importance of having a train-test split that preserves the distribution of classes.

Finally, these MAE values should be compared with the ones obtained by training our models for a pure regression problem (see Table~\ref{tab:regression_mae}). We observe that the results are comparable, and for the fridge they are even better in the reconstructed case.  The explanation comes from the fact that the raw power signal for the fridge is  almost binary, so the reconstructed signal matches this behaviour properly and the ON/OFF values $\bar{P}_{ON}$ and $\bar{P}_{OFF}$ calculated over the training set are very close to the real values. Thus, in this case good metrics for the classification problem immediately translate into good MAE for the reconstructed series. By contrast, addressing the regression problem is harder for the fridge, where the regression curve often fails to reconstruct this signal, specially when it is masked by larger signals coming from other devices (see Figure~\ref{fig:regression}b).

\subsection{Classification metrics on the regression problem}\label{sec:reg_to_class}

In the last section we trained the model for classification but evaluated the MAE for the reconstructed power signal and compared it with the MAE obtained by directly training the model for regression. Now we do the opposite: we apply the thresholding methods on the real and predicted power signal to obtain the real and predicted status at each time interval, and then we calculate the $F_1$ metric over these values. The results can be seen in Table~\ref{tab:regression_f1}.

\begin{table}[ht]
    \centering
    \begin{tabular}{|c|c|c|c|c|}
    \hline
        \textbf{Threshold} & \textbf{Model} & \textbf{Dishwasher} & \textbf{Fridge} & \textbf{Washing Machine} \\ \hline
        \multirow{2}{*}{MP} & CONV & 0.89 & 0.77 & 0.93 \\ \cline{2-5}
         & GRU & 0.90 & 0.74 & 0.91 \\ \hline
         \multirow{2}{*}{VS} & CONV & 0.77 & 0.78 & 0.81 \\ \cline{2-5}
         & GRU & 0.81 & 0.75 & 0.83 \\ \hline
         \multirow{2}{*}{AT} & CONV & 0.09 & 0.76 & 0.46 \\ \cline{2-5}
         & GRU & 0.27 & 0.72 & 0.65 \\ \hline
    \end{tabular}
    \caption{$F_1$-scores for regression models after thresholding, on each appliance and threshold}
    \label{tab:regression_f1}
\end{table}

These scores are on average worse than the ones from the original classification approach (Table~\ref{tab:classification_f1}). In particular, $F_1$-scores of AT for dishwasher and washing machine are extremely low, which is caused by the small power threshold set by the thresholding formulation (see Table~\ref{tab:activation_time}). During periods of inactivity, regression models output values that, although being relatively small compared to the power peaks of dishwasher and washing machine, are high enough to surpass the AT power threshold, thus triggering the ON state and causing many FPs.

%%%%%%%%%%%%%%%%%%%%%%%%%%%%%%%%
\section{Balancing classification and regression}\label{sec:weights}

So far we have trained our models to solve a pure classification or regression problem. However, as we explained in Section~\ref{sec:models}, our network architectures contain two different output layers, one for regression and the other for classification. This means that we can also train the model to solve both problems simultaneously, with a weighted combined loss function. 
The total loss of the model would then be:
\begin{equation}
    \mathcal{L}^{(\ell)}_{\rm tot} = w \cdot \mathcal{L}^{(\ell)}_{\rm class}  +  (1-w)\cdot \mathcal{L}^{(\ell)}_{\rm reg} / k,
\end{equation}
where $\mathcal{L}^{(\ell)}_{\rm class}$ is the binary cross entropy \eqref{eq:classloss}, $\mathcal{L}^{(\ell)}_{\rm reg}$ is the MSE \eqref{eq:regloss}, and $k$ is a constant to normalize both losses so that they have comparable magnitude, estimated to be $k = 0.0066$. The constant $w\in[0,1]$ allows to shift between pure classification and regression. 

We train both models using different values of the weight $w$ with MP thresholding. Other thresholding methods showed a similar behaviour. For each value of $w$, the model is trained five times with random weight initializations, as explained in Section~\ref{sec:methodology}. We show the output metrics MAE and $F_1$-score for varying $w$ in the figures below. 

Note that when $w=0$ the model does not train for classification. For this reason, we include for $w=0$ a single point for the $F_1$ curve, corresponding to applying the thresholding method on the regression output, as we did in Section~\ref{sec:reg_to_class}. Likewise, for $w=1$ the model does not train for regression. For this reason, we include for $w=1$ the MAE obtained by reconstructing the power signal from the classification output, as explained in Section~\ref{sec:recon}.

\begin{figure}[ht]
     \centering
     \begin{subfigure}[b]{0.48\linewidth}
         \centering
         \includegraphics[width=\textwidth]{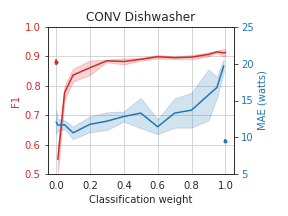}
     \end{subfigure}
     \hfill
     \begin{subfigure}[b]{0.48\linewidth}
         \centering
         \includegraphics[width=\textwidth]{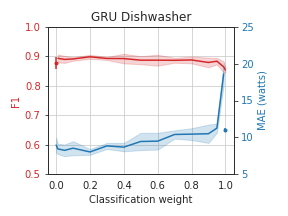}
     \end{subfigure}
     \caption{Model scores for dishwasher depending on classification weight}
     \label{fig:weight_dishwasher}
\end{figure}

\begin{figure}[ht]
     \centering
     \begin{subfigure}[b]{0.48\linewidth}
         \centering
         \includegraphics[width=\textwidth]{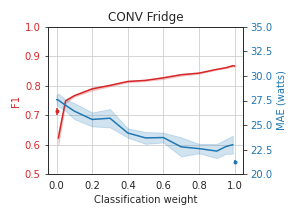}
     \end{subfigure}
     \hfill
     \begin{subfigure}[b]{0.48\linewidth}
         \centering
         \includegraphics[width=\textwidth]{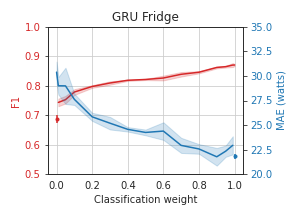}
     \end{subfigure}
     \caption{Model scores for fridge depending on classification weight}
     \label{fig:weight_fridge}
\end{figure}

\begin{figure}[ht]
     \centering
     \begin{subfigure}[b]{0.48\linewidth}
         \centering
         \includegraphics[width=\textwidth]{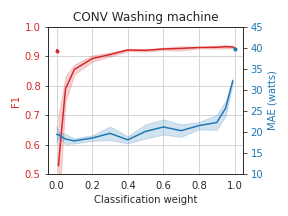}
     \end{subfigure}
     \hfill
     \begin{subfigure}[b]{0.48\linewidth}
         \centering
         \includegraphics[width=\textwidth]{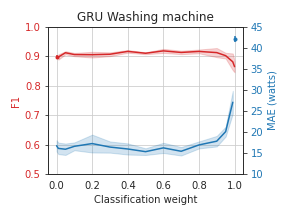}
     \end{subfigure}
     \caption{Model scores for washing machine depending on classification weight}
     \label{fig:weight_washing}
\end{figure}

Looking at the results in Figures~\ref{fig:weight_dishwasher}--\ref{fig:weight_washing} we observe a very different behaviour for the fridge than for the other two devices, due to their different characteristics already mentioned. In the dishwasher and washing machine, the $F_1$-score grows monotonically with $w$ for CONV as one would expect, but is almost constant for GRU, with a small drop for values of $w$ close to 1 which are hard to explain. Likewise, the MAE in both models and devices tends to grow for larger $w$, which is natural since the model has a smaller weight for the regression problem.
As for the extra points in the graphs, for the dishwasher we see that the $F_1$-score obtained by thresholding a pure regression output (red dot) is comparable to the best score obtained by larger weights in classification. Rather surprisingly, the reconstructed power signal from pure classification output (blue dot) has a lower MAE than the best regression weight for CONV and still a comparable value in GRU.
This discussion also holds for the washing machine, except the reconstructed series behaves poorly in this case, as we have already discussed by the activation pattern of the signal (see Table~\ref{tab:regression_reconstructed} and the ensuing remarks).
For the fridge, the behaviour is consistent for both models, but different from the other two devices. While the $F_1$-score behaves similarly, the MAE for regression decreases with $w$, which is clearly counter-intuitive: the model prioritizes classification loss, and in doing so it performs better in the regression problem as well. Also, the purely reconstructed signal for $w=1$ (blue dot) has a better MAE than any of the models trained for regression. This fact can again be explained by the second graph in Figure~\ref{fig:regression}: the regular (almost binary) activation pattern of the fridge is much better captured by a classification model with the right threshold than by a regression model, since the weak signal of the fridge is often masked by that of other devices.

%%%%%%%%%%%%%%%%%%%%%%%%%%%%%%%%

%%%%%%%%%%%%%%%%%%%%%%%%%%%%%%%%
\section{Summary and conclusions}\label{sec:outlook}
%%%%%%%%%%%%%%%%%%%%%%%%%%%%%%%%%%%%%%

Non Intrusive Load Monitoring is typically framed as a classification problem, where the input data is the aggregated power load of the household and the output data is the sequence of ON/OFF states of a given monitored device. It is important to stress that this problem is derived, as the raw data do not contain the variable that needs to be predicted but only the power consumption. Creating a classification problem from the raw power signal data requires an external determination of the status by some thresholding method. We have discussed three possible methods in Section~\ref{sec:threshold} and how they lead to  classification problems with different results.

A discussion of what is the most appropriate method should not be
based on the performance achieved by prediction models alone, but include also some objective way to judge the interpretability of the results. We suggest as an objective criterion to use the intrinsic error, i.e.  MAE  between the original power series and  reconstructed binary series. 

 We show that deep learning models can be trained to minimize the regression loss \eqref{eq:regloss} or the classification loss \eqref{eq:classloss}, but it is also possible to combine both into a weighted loss introducing an extra hyperparameter. This parameter balances the weight given to both problems, that are effectively solved both at a time. The optimal choice of this parameter depends strongly on the characteristics of the device.

To conclude we would like to mention possible improvements and future extensions of this work. 

First, two of the thresholding methods (MP and VS) are entirely algorithmic, but AT needs some external parameters to be fixed. We suggest that these free parameters, the time thresholds $(\mu_0^{(\ell)},\mu_1^{(\ell)})$ for each device, should be fine tuned to minimize the intrinsic error defined in Section~\ref{sec:recon}.

Second, some of the results highlight the importance of discussing chronological vs. random splitting of records to form the training, validation and test sets. This choice will become less significant in the large data size limit, but for moderate sizes they can still lead to different results. Similar issues are key in discussing fraud detection methods, where fraud techniques evolve in time and differ chronologically throughout the time span of the dataset.

% Generalize to unseen buildings is possible, but generalize to unseen devices is very far from the current scope of NILM research!

Finally, we have chosen to address the simpler NILM problem of recognising the same devices seen in the training and test sets. Training models on certain households and generalizing to unseen devices in different households is a harder problem that lies at the root of industrial large scale applications of NILM. 

Our purpose for this paper was to highlight an important factor at the foundations of NILM as a supervised learning problem, so we focused on a simple, well known benchmark dataset.
We envisage to extend our study to larger, more recent datasets like Pecan Street~\cite{parson2015dataport} or ECO~\cite{kleiminger2015eco}, addressing also the generalization capacity of deep learning models to cope with unseen devices in households not present in the training set.

\section{Acknowledgements}\label{sec:ack}

This research has been financed in part by the Spanish MICINN under grants PGC2018-096504-B-C33 and RTI2018-100754-B-I00 and the European Union under the 2014-2020 ERDF Operational Programme and by the Department of Economy, Knowledge, Business and University of the Regional Government of Andalusia (project FEDER-UCA18-108393).
DP gratefully acknowledges an Industrial PhD grant from Universidad de C\'adiz.

% Ensure no figure enters the references section
%\clearpage

% References
%\addcontentsline{toc}{section}{References}
%\renewcommand\refname{References}
%\printbibliography

%\bibliographystyle{plain}
\bibliographystyle{unsrt}
\bibliography{mybibfile}

\end{document}